\documentclass[epj]{svjour}
\usepackage{graphics}

\usepackage{amsmath}
\usepackage{amssymb}
\usepackage{latexsym}
\usepackage{color}
\usepackage{bm}
\usepackage{multirow}
\usepackage{cancel}
\usepackage{hyperref}

\def\beq{\begin{eqnarray}}
\def\eeq{\end{eqnarray}}
\def\beq{\begin{equation}}
\def\eeq{\end{equation}}
\def\be{\begin{eqnarray}}
\def\ed{\end{eqnarray}}

\newcommand{\lam}{\lambda}

\def\gaa{\mathrel{\raise.3ex\hbox{$>$\kern-.75em\lower1ex\hbox{$\sim$}}}}
\def\la{\mathrel{\raise.3ex\hbox{$<$\kern-.75em\lower1ex\hbox{$\sim$}}}}
\newcommand{\ba}{\begin{array}}
\newcommand{\ea}{\end{array}}
\newcommand{\besub}{\begin{subequations}}
\newcommand{\eesub}{\end{subequations}}
\newcommand{\ee}{\end{equation}}
\newcommand{\bea}{\begin{eqnarray}}
\newcommand{\eea}{\end{eqnarray}}

\begin{document}
\title{Bosonic Decays of Charged Higgs Bosons in a 2HDM Type-I}
\author{A.~Arhrib\inst{1} \and R.~Benbrik \inst{2} \and S.~Moretti \inst{3} 
}                     
%
%
\institute{Facult\'e des Sciences et Techniques, Abdelmalek Essaadi University, B.P. 416, Tangier, Morocco.\\
Physics Division, National Center for Theoretical Sciences, Hsinchu, Taiwan.
 \and Physics Division, National Center for Theoretical Sciences, Hsinchu, Taiwan.\\ 
LPHEA, Semlalia, Cadi Ayyad University, Marrakech, Morocco.\\ MSISM Team, Facult\'e Polydisciplinaire de Safi, 
Sidi Bouzid, B.P. 4162,  Safi, Morocco.
\and School of Physics and Astronomy, University of Southampton,
  Southampton, SO17 1BJ, United Kingdom.
}

%
%
\abstract{
In this study, we focus on the bosonic decays of light charged Higgs
  bosons in the 2-Higgs Doublet Model (2HDM) Type-I. We quantify the Branching Ratios (BRs) of the $H^\pm \to W^\pm
  h$ and $H^\pm\to W^\pm A$ channels and show that they could be substantial over several areas
  of the parameter space of the 2HDM Type-I that are  still allowed by Large Hadron Collider (LHC) and other experimental
 data as well as 
  theoretical constraints.  We suggest that $H^\pm \to W^\pm h$ and/or $H^\pm \to W^\pm A$ could be
  used as a feasible discovery channel alternative to $H^\pm \to \tau\nu$.
%
} 
\maketitle
\section{Introduction}
\label{intro}
Following the discovery of a 125 GeV Higgs boson in the first 
run of the LHC \cite{Aad:2012tfa,Chatrchyan:2012xdj}, 
several studies of its properties were undertaken. The 
current situation is that 
the measured Higgs
signal rates in all channels agree  with the Standard 
Model (SM) predictions at the
$\sim2\sigma$ level~\cite{atlasandcms}. 
Although the current LHC Higgs data are consistent with the SM, 
there is still the possibility that the
observed Higgs state could be part of a model with an extended 
Higgs sector including, e.g., an  extra doublet, singlet and/or triplet. 
As the discovered Higgs state belongs to a doublet, we concern 
ourselves here with such a scenario.
Most of the higher Higgs representations with an extra doublet
predict in their spectrum one or more charged Higgs bosons. Discovery of
such a state would therefore be an indisputable signal of an extended Higgs
sector and a clear evidence for a departure from the SM.  
One of the main goals of the 13 TeV LHC
(eventually to be upgraded to 14 TeV) is  to improve the 
precision of the  measurements of the 
Higgs couplings, thus to access potential new physics 
indirectly. However, in parallel, direct searches for new 
Higgs states
will also take place in the quest to  find an evidence of 
physics Beyond the SM (BSM).
One of the simplest extensions of the SM is the 2HDM, which contains
 two Higgs doublets, $H_1$ and $H_2$, used to give mass to all 
 fermions. The particle spectrum of the 2HDM
 is as follows: two CP even ($h$ and $H$, with $m_h<m_H$), 
one CP odd ($A$) and a pair of charged  ($H^\pm$) Higgs
bosons. At hadron colliders, a charged Higgs boson
can be produced through several channels. Light charged Higgs states, i.e, 
with $m_{H\pm}\leq m_t-m_b$, are copiously induced by  $t\bar{t}$ production
followed by the top decay $t\to bH^+$ (or the equivalent antitop mode).
When kinematically allowed, 
$pp\to t\bar{t}\to b\bar{b}H^- W^+ $  + c.c. provides the most 
important source of light charged Higgs bosons, above and beyond the yield of 
 various direct production modes:
$gb \to tH^-$ and $gg\to t\bar{b}H^-$ \cite{single},
$gg \to W^\pm H^\mp $ and
$b\bar{b} \to  W^\pm H^\mp $ \cite{wh},
$q\bar q' \to \phi H^\pm$ where 
$\phi$ denotes one of the  three neutral 
 Higgs bosons~\cite{cpyuan},
$gg\to H^+H^-$ and $q\bar q\to H^+H^-$ 
 \cite{hadronic},
 $qb\to q' H^+ b$ \cite{moar} and
 $c \bar s, c\bar b\to H^+$ 
\cite{Dittmaier:2007uw}. (See also Refs.~\cite{Aoki:2011wd,Akeroyd:2016ymd} for a review of all available $H^\pm$ hadro-production modes
in 2HDMs.)

At the Tevatron and LHC, light charged Higgs bosons can be detected through $pp\to t\bar{t}\to b \bar{b}H^- W^+$ 
 followed by $\tau\nu$ decay. In fact, for a  light charged Higgs state,
the $\tau\nu$ decay is  the dominant  mode. 
The ATLAS and CMS experiments have already drawn  an exclusion on BR$(t\to bH^+) \times
{\rm BR}(H^\pm \to \tau \nu)$ based on the search for the corresponding decay chain 
\cite{Aad:2014kga,Khachatryan:2015qxa}. Other channels, such as  
$H^+ \to c\bar{s}$, have also been searched for by ATLAS and CMS 
\cite{Aad:2013hla,Khachatryan:2015uua}. Assuming that BR$(H^+ \to
c\bar{s})=100\%$, one can set a limit on BR$(t\to bH^+)$ to be in the range
$5\%$ to $1\%$ for a charged Higgs mass between 90 and 150 GeV.
We remind here in passing that charged Higgs bosons have been also searched for at LEP-II 
using charged Higgs pair production followed by either $H^\pm \to \tau \nu$, 
$H^\pm \to c{s}$ or $H^\pm \to W^\pm A$
\cite{Abbiendi:2013hk}. If the charged Higgs boson decays 
dominantly to $\tau\nu$ or $c{s}$, the LEP-II
lower bound on the mass is of the order of 80 GeV while in the case where charged Higgs 
decay is dominated by $W^{\pm *}A$, via a  light CP-odd Higgs state ($m_{A}\approx 12$ GeV), the lower
bound on the charged Higgs mass is about 72 GeV \cite{Abbiendi:2013hk}.

The aim of this letter is  to show that the bosonic decays of a {\sl light} 
charged Higgs boson, such as $H^\pm \to W^{\pm *}h$ and/or  
$H^\pm \to W^{\pm *}A$, could be substantial and may compete with  
$H^\pm\to \tau\nu$ and $ c{s}$\footnote{These channels have been 
studied previously in \cite{Akeroyd:1998dt}. 
We show here that this possibility is consistent with LHC data.}.
In particular, 
$H^\pm \to W^{\pm *}h$ with leptonic decay of $W^\pm$ could be an alternative channel to 
$H^\pm\to \tau\nu$ in order to discover a light charged Higgs boson at the LHC
owing to the handle offered by the SM-like Higgs mass reconstruction, now possible after discovery \cite{Enberg:2015qsa}--\cite{Enberg:2014pua}.
We also discuss the case of a light CP-odd Higgs $m_A\leq 120$ GeV
 where $H^\pm \to W^{\pm *} A$ could be substantial and reach a 100\% branching
 fraction while  being consistent with LHC and LEP data. This possibility may
 suggest that a light charged Higgs state could  have escaped detection during  previous LHC
 searches. Therefore, the bosonic decays of a 2HDM charged Higgs boson might be
 complementary to the usual search channels $H^\pm\to \tau\nu$ and $H^\pm\to cs$.
We also point out that $H^\pm \to W^{\pm *}h/A$ would lead to the same final state as
$H^\pm \to t^*b\to W^* b\bar{b}$ in the case where $h$ and $A$ decay to
$b\bar{b}$. Clearly, there are kinematic differences between these three channels,
so that one can eventually separate them, e.g.,  by reconstructing the
$b\bar b$ pair around a Higgs resonance (125 GeV or others) and/or the $bW^\pm$ pair around the 
(anti)top pole, yet it may pay off to devise an inclusive approach that maximizes the signal yield across the three
decay decay patterns \cite{Moretti:2016jkp}.
\section{A review of the 2HDM}
\label{sec:1}
The most general renormalizable potential  for a model of exactly
two scalar Electro-Weak (EW) doublets with the quantum numbers 
which are invariant under $SU(2)\otimes U(1)$ can be written as
\begin{eqnarray}
\label{higgspot}
V(\Phi_1,\Phi_2) &=& m^2_1 \Phi^{\dagger}_1\Phi_1+m^2_2
\Phi^{\dagger}_2\Phi_2 + (m^2_{12} \Phi^{\dagger}_1\Phi_2+{\rm
h.c}) \\\nonumber &+&\frac{1}{2} \lam_1 (\Phi^{\dagger}_1\Phi_1)^2 +\frac{1}{2}
\lam_2 (\Phi^{\dagger}_2\Phi_2)^2\nonumber \\\nonumber &+& \lam_3
(\Phi^{\dagger}_1\Phi_1)(\Phi^{\dagger}_2\Phi_2) + \lam_4
(\Phi^{\dagger}_1\Phi_2)(\Phi^{\dagger}_2\Phi_1) \\\nonumber &+& \frac{1}{2}
\lam_5[(\Phi^{\dagger}_1\Phi_2)^2+{\rm h.c.}] , 
\end{eqnarray}
where $\Phi_i$, $i=1,2$ are complex $SU(2)$ doublets with 4
degrees of freedom each and  $m_{i}^2$, $\lambda_i$ and $m_{12}^2$ are real
which follows from the hermiticity of the potential. From the initial 8
degrees of freedom, if the $SU(2)$ symmetry is broken, we end up with the aforemetioned 5 physical Higgs states, upon the absorption  of 3 Goldstone bosons by the $W^\pm$ and $Z$ states.
The potential in Eq.~(\ref{higgspot}) has a total of 10 parameters if one includes the vacuum expectation values. In a CP-conserving minimum there are two minimization conditions that can be used to fix the tree-level value of the parameters $m_1^2$ and $m_2^2$. The combination $v^2=v_1^2 + v_2^2$ is fixed as usual by the EW breaking scale through $v^2=(2\sqrt{2} G_F)^{-1}$.  We are thus left with 7 independent parameters, namely $(\lambda_i)_{i=1,\ldots,5}$, $m_{12}^2$, and $\tan\beta \equiv v_2/v_1$.  Equivalently, we can take instead the set $m_{h}$, $m_{H}$, $m_{A}$, $m_{H^\pm}$, $\tan\beta$, 
$\sin(\alpha -\beta)$ and $m_{12}^2$ as the 7 independent parameters. The angle $\beta$ is the rotation angle from the group eigenstates to the mass eigenstates in the CP-odd and charged sector. The angle $\alpha$ is the corresponding rotation angle for the CP-even sector. The parameter $m_{12}$ is a measure of how the discrete symmetry is broken. The potential with $m_{12}=0$ has an exact $Z_2$ symmetry and is always CP-conserving.

\begin{figure*}[t!]
\resizebox{0.5\textwidth}{!}{
\includegraphics{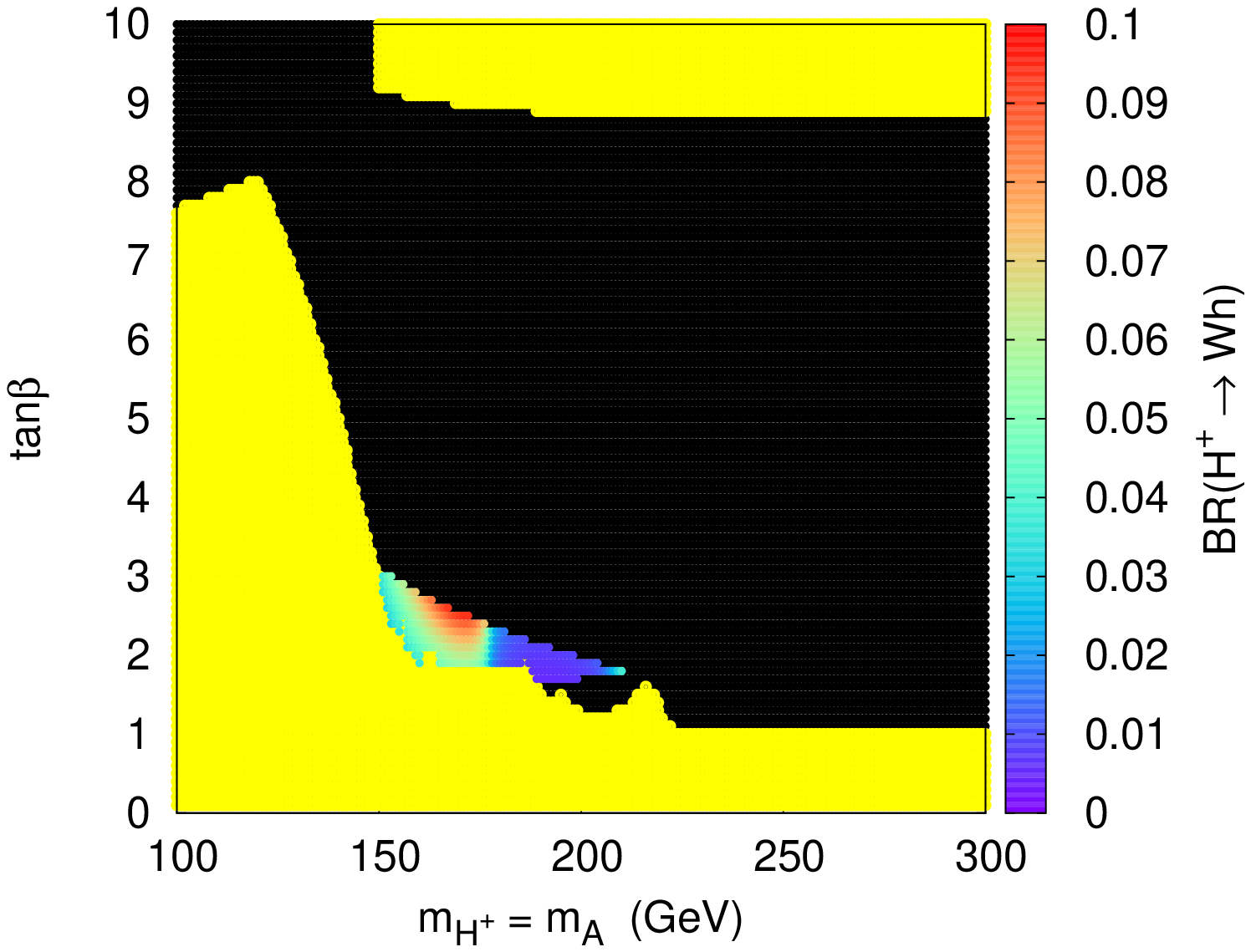}}
\resizebox{0.5\textwidth}{!}{
\includegraphics{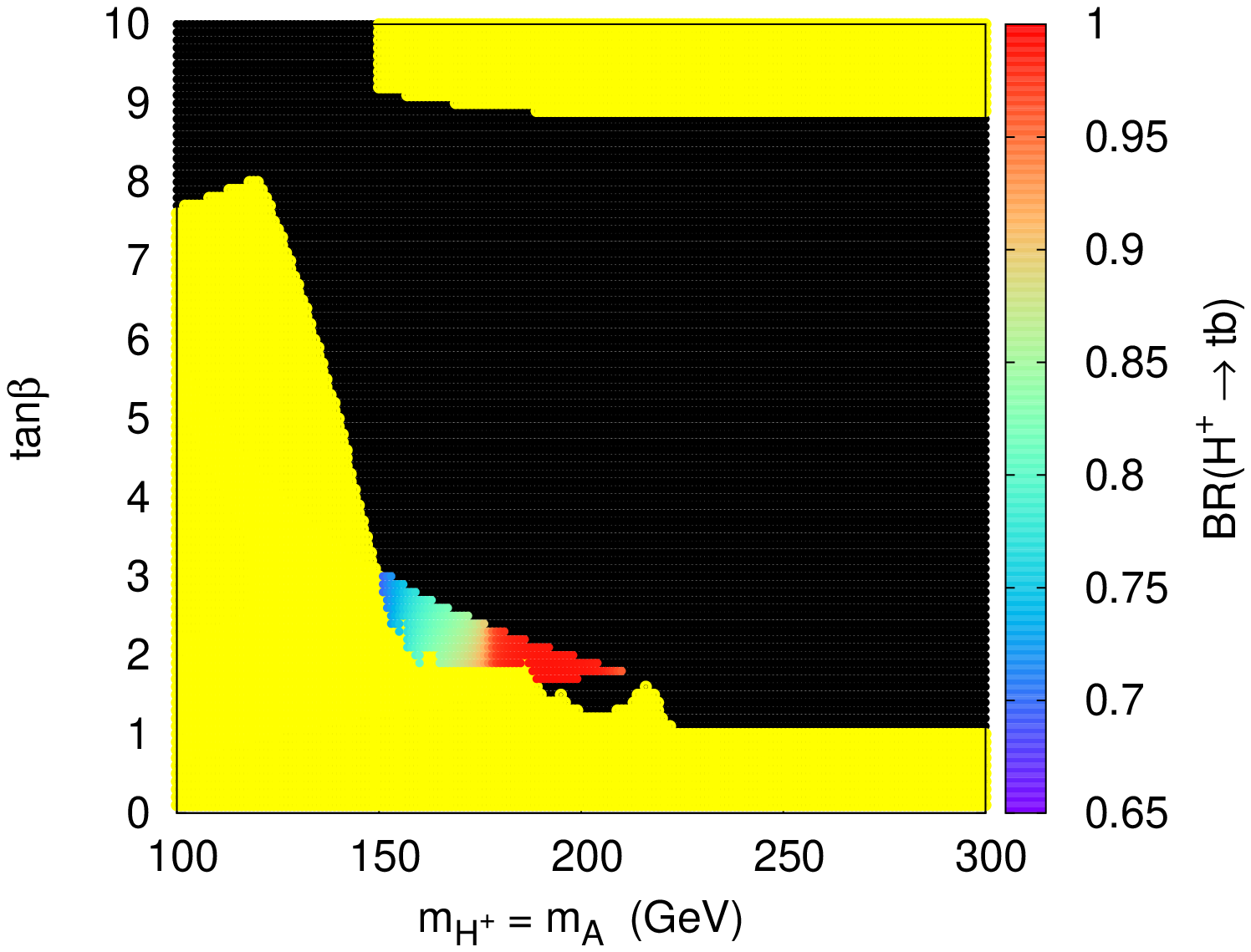}}
\caption{The  BR$(H^\pm \to W^{\pm *} h)$  (left) and BR$(H^\pm \to t^*b)$  (right)
   in the 2HDM-I mapped over the $(m_{H^\pm},\tan\beta)$ plane with 
$m_{H^\pm} = m_A $, $m_H = 300$ GeV and $\sin(\beta - \alpha) = 0.85$. We set $m^2_{12} = m^2_{H^\pm} s_\beta c_\beta$.
 Yellow color areas are excluded
   from LHC Higgs data at 95\% Confidence Level (CL) while black/grey ones are excluded from
   theoretical constraints.
}
\label{fig:mctb1}
\end{figure*}

\section{Theoretical and experimental bounds}
\label{sec:bounds}
The parameter space of the scalar potential of the 2HDM is reduced both by theoretical constraints as well as by the results of experimental searches. Amongst the theoretical constraints which the 2HDM is subjected to, we start by requiring vacuum stability of the theory. We also force the potential to be perturbative by requiring that all quartic couplings of the scalar potential, Eq.~(\ref{higgspot}), obey $|\lambda_i| \leq 8 \pi$ for all $i$. For the vacuum stability conditions, that ensure that the potential is bounded from below, we use those from \cite{vac1}, which are given by
\begin{eqnarray}
\nonumber
& \lambda_1  > 0, \lambda_2 > 0,
& \sqrt{\lambda_1\lambda_2 }
+ \lambda_{3}  + {\rm{min}}
\left( 0 , \lambda_{4}-|\lambda_{5}|
 \right) >0. \label{vac}
\end{eqnarray}

However, the most restrictive theoretical bounds come from the full set of unitarity constraints \cite{unit1,abdesunit} established using the high energy approximation as well as the equivalence theorem and which can be written as
\begin{eqnarray}
  |a_{\pm}|,  |b_{\pm}|,  |c_{\pm}|,  |d_{\pm}|,
   |e_{1,2}^{}|,  |f_{\pm}|,  |g_{1,2}^{}|  < 8 \pi \label{unita}
\end{eqnarray}
with
\begin{eqnarray}
a_{\pm}^{} &=&
 \frac{3}{2} \left\{
  (\lambda_1 + \lambda_2) \pm
   \sqrt{ (\lambda_1-\lambda_2)^2 + \frac{4}{9} (2\lambda_3+\lambda_4)^2}
  \right\},  \\
b_{\pm}^{} &=&
 \frac{1}{2} \left\{
   (\lambda_1 + \lambda_2) \pm
   \sqrt{ (\lambda_1-\lambda_2)^2 +4 \lambda_4^2}
  \right\},  \\
c_{\pm}^{} &=& d_{\pm}^{}=
 \frac{1}{2} \left\{
   (\lambda_1 + \lambda_2) \pm
   \sqrt{(\lambda_1-\lambda_2)^2 +4 \lambda_5^2}
  \right\},  \\
e_1 &=&    \left(
    \lambda_3 + 2 \lambda_4 - 3 \lambda_5 \right), \qquad  \qquad
e_2 =    \left(
    \lambda_3 - \lambda_5 \right), \\
f_+ &=&    \left(
    \lambda_3 + 2 \lambda_4 + 3 \lambda_5 \right), \qquad  \qquad
f_- =    \left(
    \lambda_3 + \lambda_5 \right), \\
g_1 &=& g_2 =    \left(
    \lambda_3 + \lambda_4 \right).
\end{eqnarray}
The 2HDM parameters are also constrained by direct experimental searches and by precision experimental data. First, the extra contributions to the $\delta \rho$ parameter from the extra Higgs scalars \cite{Rhoparam} should not exceed the current limits from precision measurements \cite{pdg4}: $ |\delta\rho| \la 10^{-3}$.
Values of $\tan \beta$ smaller than $\approx 1$ are disallowed both by the constraints coming from $Z \rightarrow b \bar{b}$ and from $B_q \bar{B_q}$ mixing~\cite{Oslandk} for all Yukawa versions of the model. Conversely,  $\tan \beta$ cannot be too large due to the aforementioned theoretical constraints. 
We also require agreement with the null-searches from the LEP, Tevatron and LHC experiments. Finally, we require agreement within 2$\sigma$ for the 125 GeV Higgs signal strength measurements.
\section{Discussion}
\begin{figure*}[t!]
\resizebox{0.5\textwidth}{!}{
\includegraphics{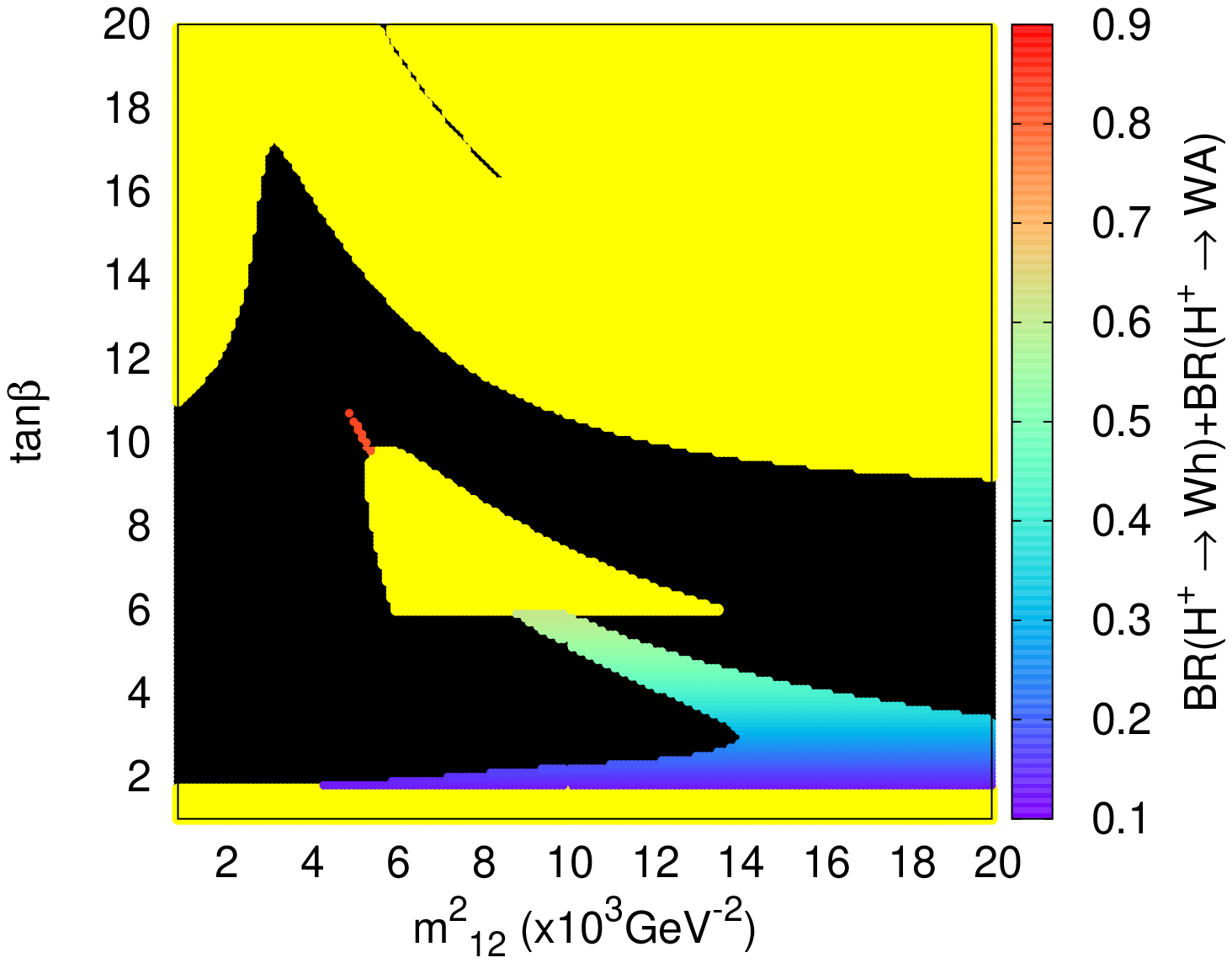}}
\resizebox{0.5\textwidth}{!}{
\includegraphics{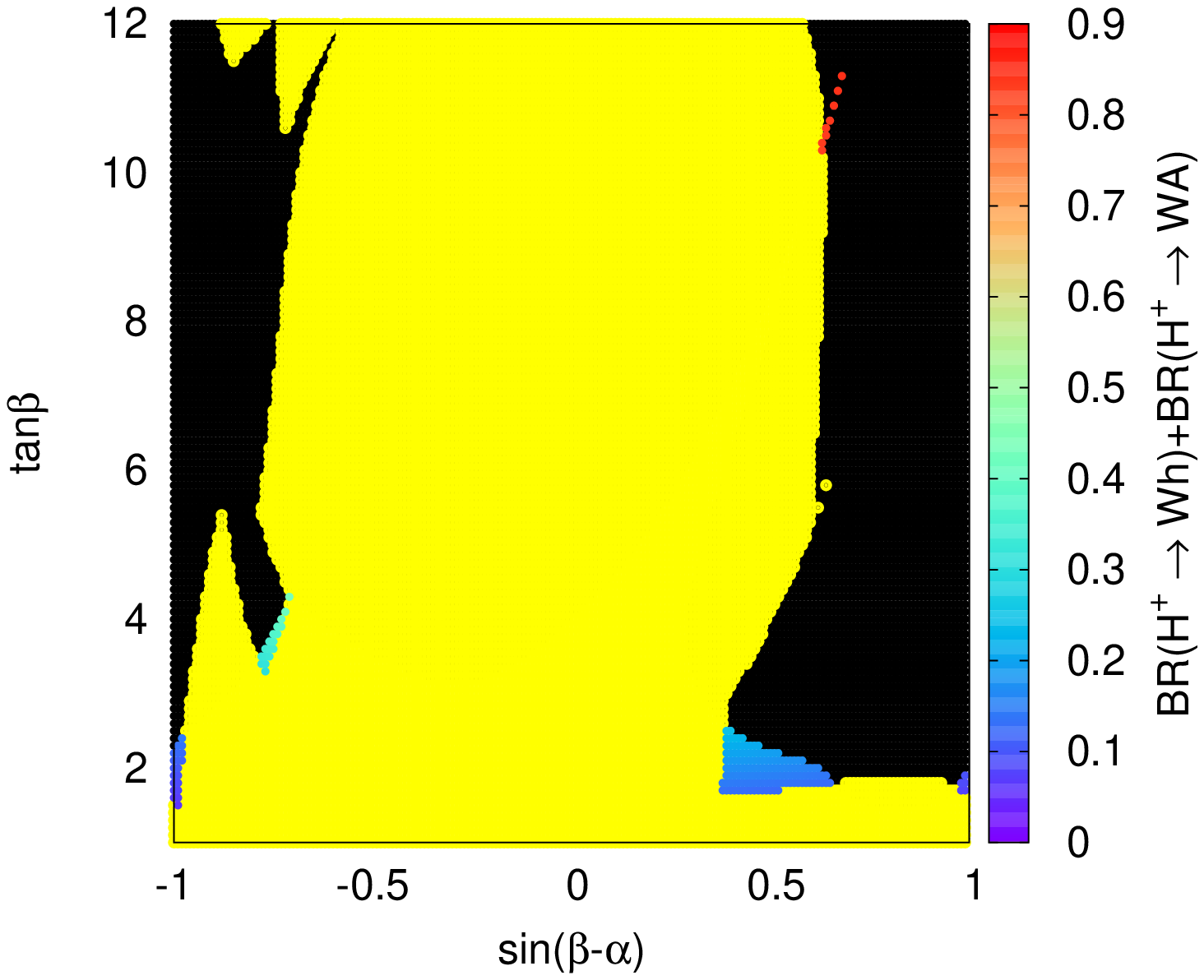}}
\caption{The BR$(H^\pm \to W^{\pm *}h+W^{\pm *}A)$ in the 2HDM-I mapped over the $(m_{12}^2,\tan\beta)$ (left) and $(\sin(\beta-\alpha),\tan\beta)$ (right) planes with  $m_A = m_h = 125 $ GeV, $m_{H^\pm} = 170$ GeV and $m_{H}  = 300$ GeV.
The other parameters are $\sin(\beta-\alpha) = 0.65$ (left) and $m^2_{12} = 5000$ GeV$^{-2}$ (right). Colour coding is the same as in Fig. \ref{fig:mctb1}.
}
\label{fig:mctb3}       
\end{figure*}
\begin{figure*}[t!]
\resizebox{0.5\textwidth}{!}{
\includegraphics{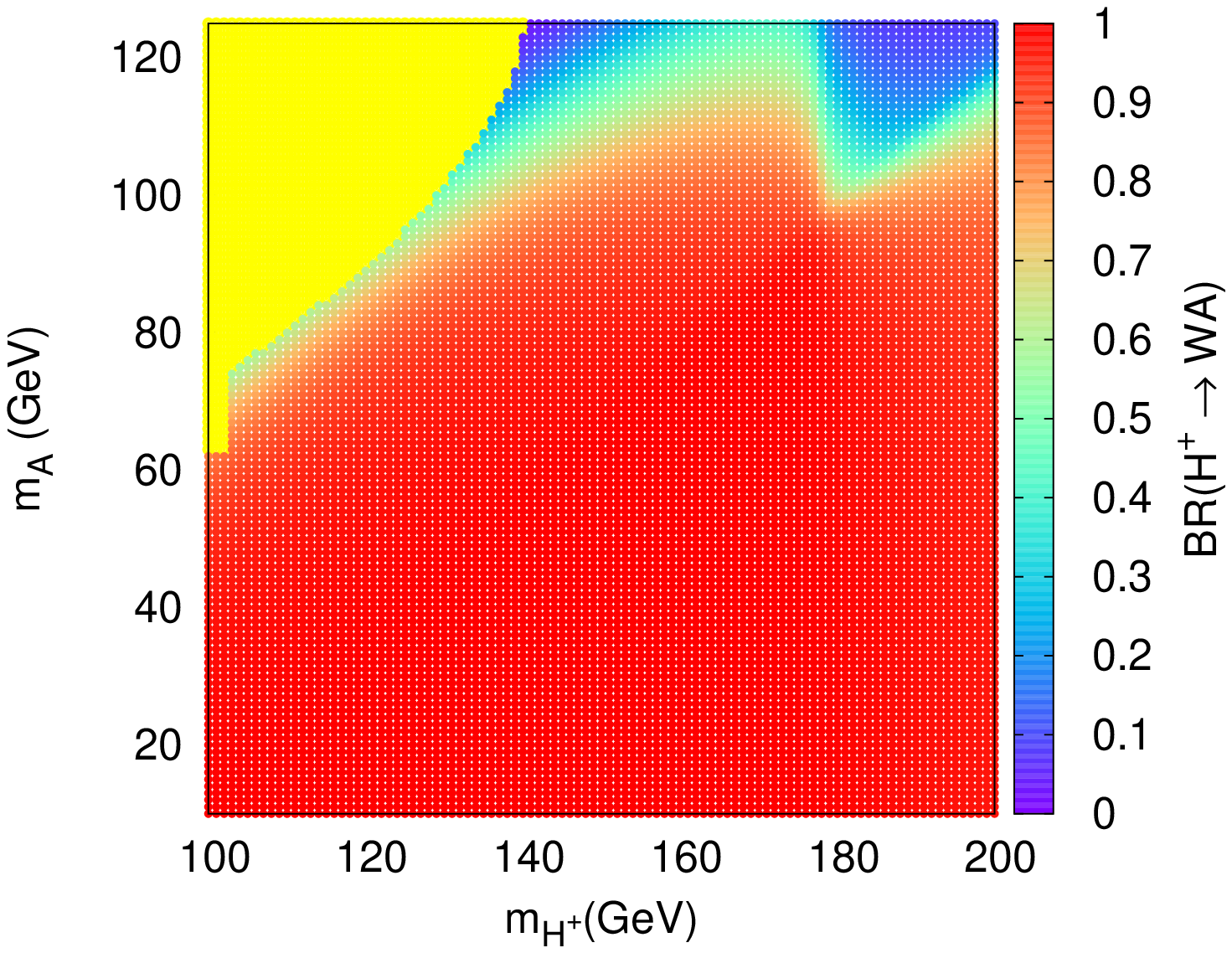}}
\resizebox{0.5\textwidth}{!}{
\includegraphics{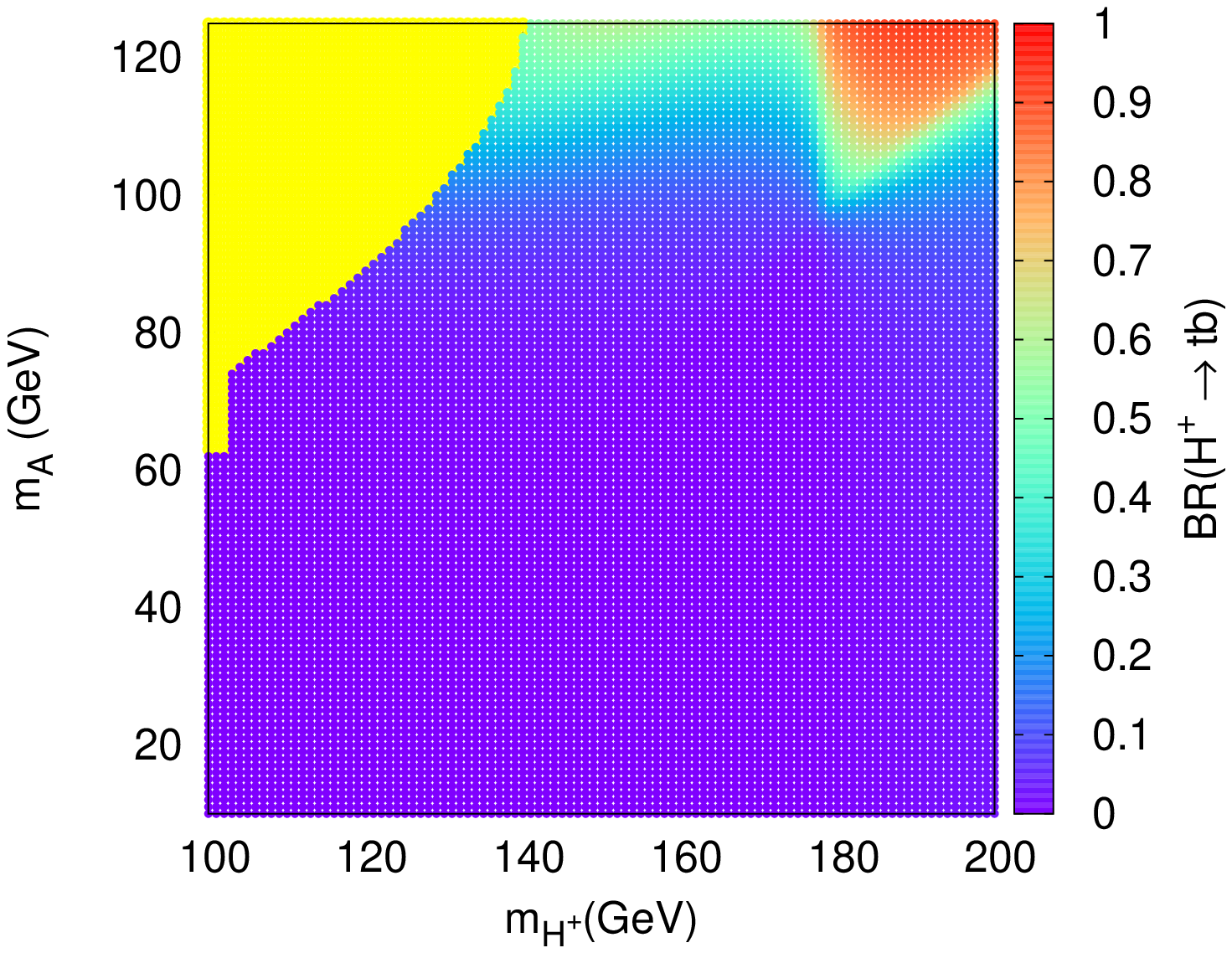}}
\caption{The BR$(H^\pm \to W^{\pm *}A)$  (left) and BR$(H^\pm \to t^*b)$  (right) rates
  in the 2HDM-I mapped over $(m_{H^\pm},m_A)$ plane for the following
  parameter choice: $m_h = 125 $ GeV, $\sin(\beta-\alpha) = 1$, $\tan\beta = 5$, $m_{H}  = 300$ GeV and $m^2_{12} = 16\times 10^3$ GeV$^{2}$. The yellow region is excluded by LHC data at $95\%$ CL.}
\label{fig:mcma2}
\end{figure*}
\begin{figure*}[t!]
\resizebox{0.5\textwidth}{!}{
\includegraphics{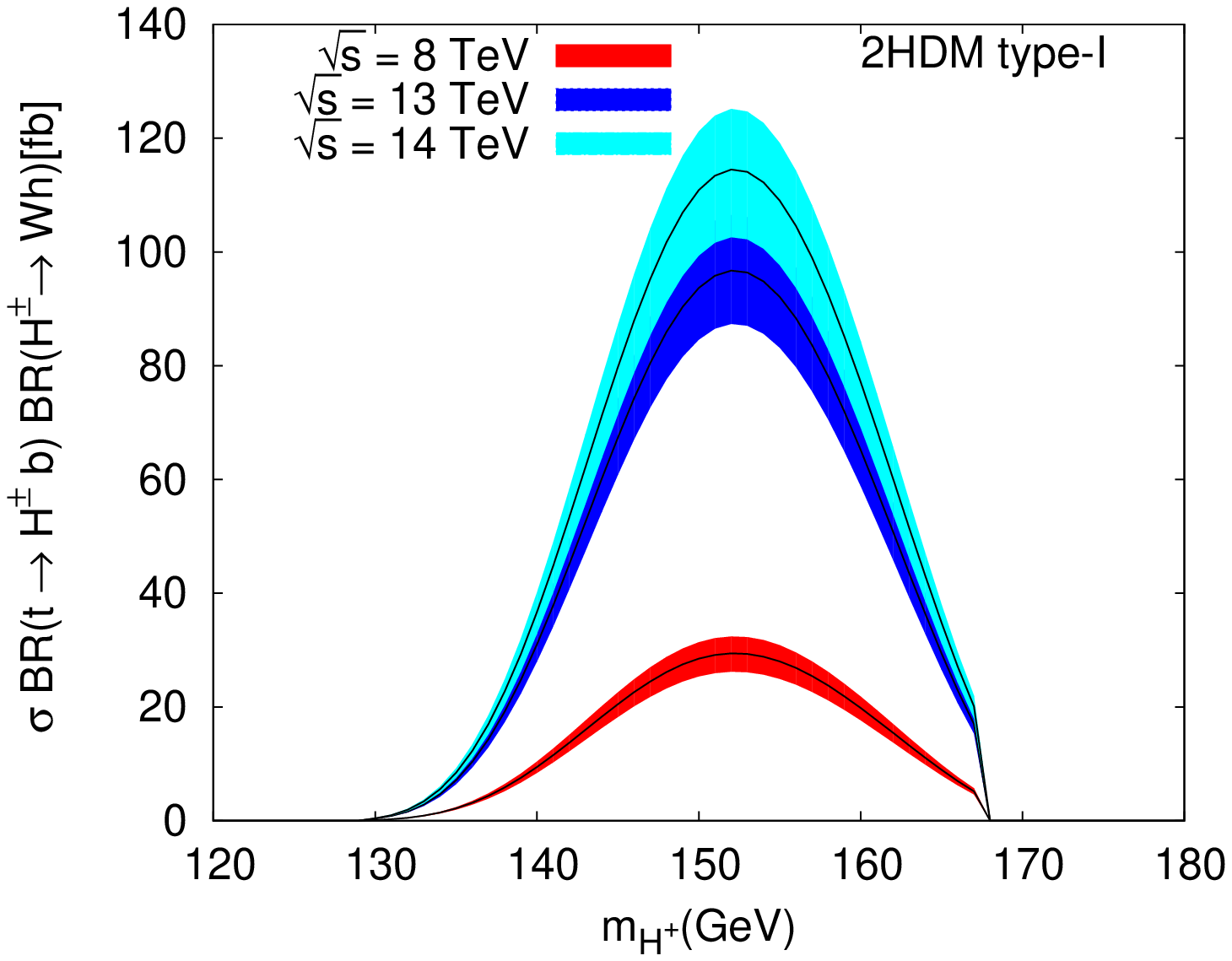}}
\resizebox{0.5\textwidth}{!}{
\includegraphics{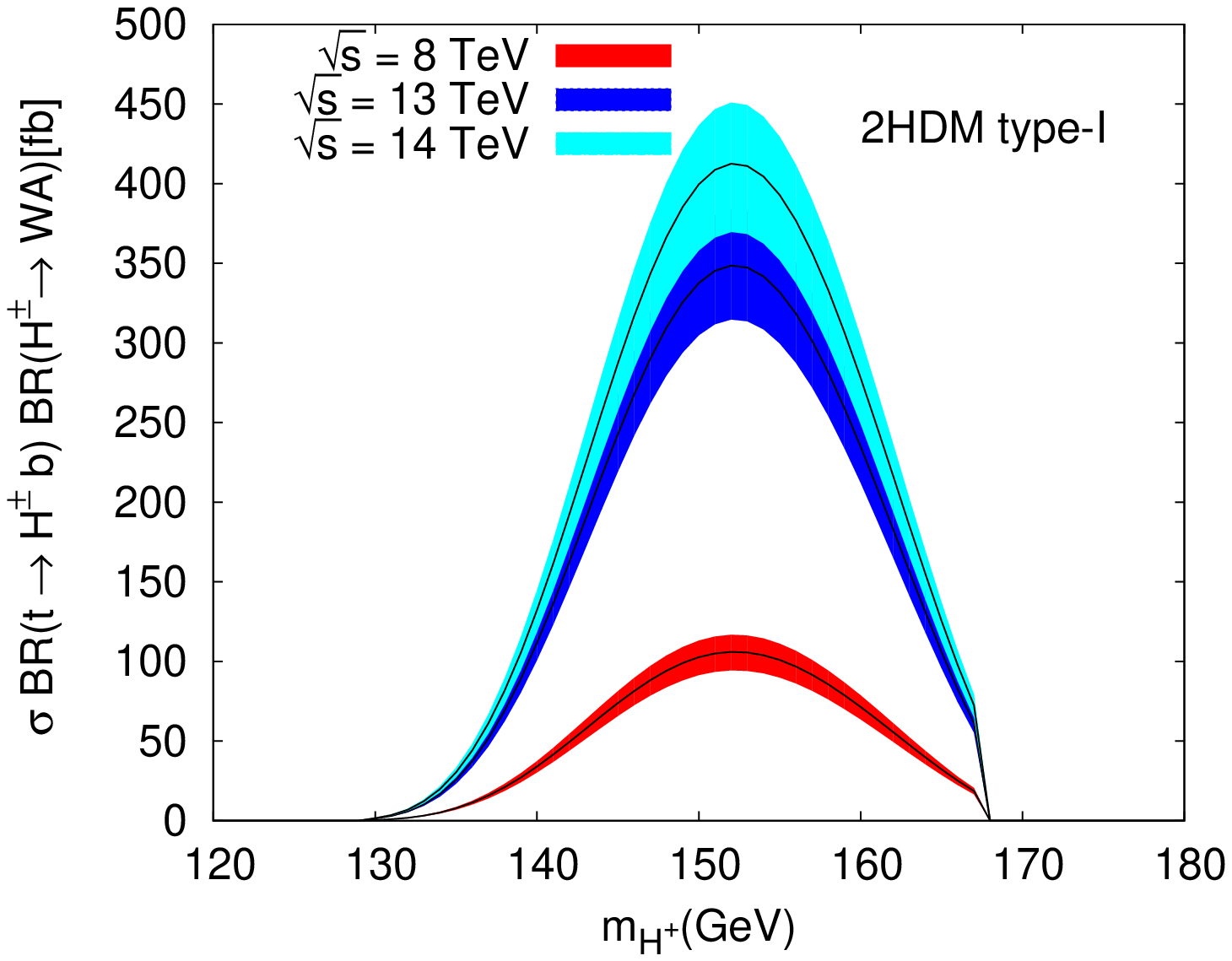}}
\caption{The rates for $\sigma(pp\to t\bar{t})\times{\rm  BR}(t\to H^\pm b)\times {\rm BR}(H^\pm\to
  W^{\pm *}\phi)$  with $\phi =h$  (left) and  $A$  (right) in the 2HDM-I as a function of
  $m_{H^\pm}$ with $m_A = m_h=125$ GeV for $\tan\beta = 5$, 
$\sin(\beta - \alpha) = 0.85$, $m_H = 300 $ GeV and 
$m^2_{12} = 16\times 10^3$ GeV$^{2}$. The bands correspond to a 1$\sigma$
 deviation from the $t\bar{t}$ cross section central value as computed at Next-to-Next-to-Leading Order (NNLO)
at three LHC energies.}
\label{fig:mctb4}
\end{figure*}
It is well known that, in the framework of a 2HDM Type-II (hereafter, 2HDM-II), the $b\to s\gamma$ constraints force
the charged Higgs mass to be  heavier than 580~GeV~\cite{Misiak:2017bgg,Misiak:2015xwa} 
for any value of $\tan\beta \geq 1$. 
Therefore, in the
present study, we will deal only with a 2HDM Type-I (henceforth, 2HDM-I) where a light
charged Higgs state is still allowed by all $B$-physics constraints
\cite{Enomoto:2015wbn} {{so long that $\tan\beta\geq 1.5$.}} 

In this study, $h$ is taken to be the SM-like Higgs
 boson and will be fixed at 125 GeV. The other parameters are varied within
 a specific range in order to satisfy theoretical  as well as experimental
 constraints. We have used the public code 2HDMC-1.7.0 \cite{Eriksson:2009ws} 
to perform the scan over the 2HDM parameter space.
The program is also linked to HiggsBounds-4.3.1 and HiggSignals-1.4.0 
\cite{Bechtle:2008jh} to check against  available collider constraints.
A systematic scan is performed over $m_A$, $m_{H\pm}$, $\tan\beta$ and $\sin(\beta - \alpha)$. The mixing angle $\alpha$ is fixed from $\sin(\beta-\alpha)$. The mass of the heavy CP-even Higgs boson was fixed at $m_H = 300$ GeV.
In Fig.~\ref{fig:mctb1}, we scan over the $(m_{H^\pm} ,\tan\beta)$ plane 
and set $ m_{H^\pm} = m_A $  with $\sin(\beta - \alpha) = 0.85$ while  
$m^2_{12}$ is fixed to $m_A^2$.  
The black/grey regions are excluded from theoretical constraints while 
 the yellow region is excluded by experimental constraints at 95\%CL. 
It is clear  that a light  charged Higgs state with mass $\leq 150$ GeV is 
excluded from $H^\pm\to \tau\nu$ and $H^\pm\to c{s}$ searches 
\cite{Aad:2014kga,Khachatryan:2015qxa,Aad:2013hla,Khachatryan:2015uua}. 
We are left with a small region with $m_{H\pm}\in [150,210]$ GeV 
 in which we have evaluated 
BR$(H^\pm \to W^\pm h)$  and BR$(H^\pm \to t^*b)$. The two BRs are quantitatively
shown in the vertical palettes: left panel is for BR$(H^\pm \to W^{\pm *} h)$  
and right panel is for BR$(H^\pm \to t^*b)$. 
One can see that, in this scenario, 
before the top-bottom threshold, BR$(H^\pm \to W^{\pm *} h)$ can reach 
10\% for a charged Higgs mass around 160 GeV
and $2\leq \tan\beta\leq 3$. After crossing the top-bottom threshold, 
BR$(H^\pm \to W^{\pm *} h)$ becomes suppressed and BR$(H^\pm \to t^*b)$ gets
enhanced so as to dominate over all other decays.

In Fig.~\ref{fig:mctb3} we  illustrate the size of 
BR$(H^\pm \to W^{\pm *}h+W^{\pm *}A)$ 
over the ($m_{12}^2,\tan\beta$) plane  (left) 
and  ($\sin(\beta-\alpha),\tan\beta$) plane  (right) for
$m_{H^\pm} = 170$ GeV, $m_{H}  = 300$ GeV, $m_{h}=m_A  =125$ GeV. 
In the left panel we 
show the effect of the soft $Z_2$ breaking term $m_{12}$. 
It is clear that
for some special $m_{12}^2$ and $\tan\beta$ choices, the 
BR$(H^\pm \to W^{\pm *}h+W^{\pm *}A)$ could reach 90\%.
In the right panel, one can see that 
LHC data favor $\sin(\beta-\alpha)$ to be rather close to the
decoupling limit: $\sin(\beta-\alpha)\approx 1$, which implies
$\cos(\beta-\alpha)\approx 0$. Therefore, the coupling $W^\pm H^\mp h$, which
is proportional to $\cos(\beta-\alpha)$, is suppressed
for $\sin(\beta-\alpha)\approx 1$ while the $W^\pm H^\mp A$ one, which is a gauge
coupling, has no suppression factor. 
This fact will make BR$(H^\pm \to W^{\pm *}A)$ larger than 
BR$(H^\pm \to W^{\pm *}h)$ in the special case of $m_h=m_A$.
In this scenario where $m_h=m_A$, BR$(H^\pm \to t^*b)$ and
BR$(H^\pm \to W^{\pm *}h+W^{\pm *}A)$ are anti-correlated, i.e., when 
BR$(H^\pm \to W^{\pm *}h+W^{\pm *}A)$ is maximal BR$(H^\pm \to t^*b)$
is suppressed and vice versa.

We now turn to the case of a light CP-odd Higgs state, 
with $m_A\leq 125$ GeV. Such a Higgs state
 is still allowed by LEP-II and LHC data. 
In Fig.~\ref{fig:mcma2} we scan over both the 
CP-odd and charged Higgs boson masses 
over the following region: $10$ GeV $\leq m_A\leq$ 120 GeV, 
$100$~GeV~$ \leq m_{H\pm} \leq$ 200 GeV with $m_h =$ 125  GeV, 
$\sin(\beta-\alpha)$ = $1$,  $\tan\beta$ $=$ 5 and $m_{H}=300$ GeV. 
In this scan, the yellow region is where $H^\pm \to W^{\pm *} A$ is kinematically
not allowed and therefore the charged Higgs boson will decay dominantly to 
$\tau\nu$ and/or $c{s}$ pairs and is excluded by LHC data. However,
over a substantial area of the $(m_{H^\pm},m_A)$ plane, it is clear
that BR$(H^\pm \to W^{\pm *}A)$ can be the dominant decay channel, 
i.e, for $m_A\leq 100$ GeV for any value of the charged Higgs mass 
and in such a case  BR$(H^\pm \to t^*b)$ becomes a sub-leading channel.

We finally show in Fig.~\ref{fig:mctb4} the single charged Higgs production cross section where the
$H^\pm$ state comes from  (anti)top decays following $t\bar{t}$ 
hadro-production: 
$\sigma(pp\to t\bar{t})\times {\rm BR}(t\to
H^\pm b)\times {\rm BR}(H^\pm\to W^{\pm *}\phi)$, where $\phi=h$ or $A$.  
We plot $\sigma$ (for $\phi=h$ on the left and for $\phi=A$ on the right) 
  as a function of the charged Higgs mass for $m_A = m_h=125$ GeV, 
  $\tan\beta = 5$, $\sin(\beta - \alpha) = 0.85$, $m_H=300$ GeV
and $m^2_{12}= 16\times 10^3$ GeV$^{2}$. Both cross sections reach their
maximum values for $m_{H\pm}\approx 150$ GeV. In the case of $\phi=A$ the
cross section is larger than in the case $\phi=h$ 
and can be of order 400--450 fb at the two highest LHC energies of 13--14 TeV. 
Notice that the former is larger than the latter primarily because BR$(H^\pm \to W^{\pm *}A)$ can be  about 4 times larger  than BR$(H^\pm \to W^{\pm *}h)$ (for the same Higgs boson masses, as discussed).
In conclusion, we have proven the existence within the 2HDM-I of sizable regions of the parameter space compliant with
all available theoretical and experimental constraints yielding substantial BRs for $H^\pm$ decays into $W^{\pm *}h$
(with $h$ being the SM-like Higgs state) and/or $W^{\pm *}A$ in which the $H^\pm$ mass is less than $m_t-m_b$,
wherein $W^{\pm *}\to l\nu$ ($l=e,\mu$).
Under the circumstances, $H^\pm$ production in single mode from the decay of a(n) (anti)top quark is possible with high rates, which are indeed potentially accessible during the present Run 2 of the LHC.  These regions of parameter space within the 2HDM-I are ameanable to immediate experimental investigation by the ATLAS and CMS collaborations, which have
so far concentrated their attention almost exclusively onto $\tau\nu$ and/or $cs$ decays of a light charged Higgs state emerging from $t\bar t$ production and decay.  \\
\vspace{-0.75cm}
\section*{Acknowledgements}
\vspace{-0.2cm}
The authors are supported by the grant H2020-MSCA-RISE-2014 no. 645722
(NonMinimalHiggs). 
This work is also supported by the Moroccan Ministry of Higher
Education and Scientific Research MESRSFC and  CNRST: Projet PPR/2015/6.
SM is supported in part through the NExT Institute. 
RB thanks SM for hospitality in Southampton when part of this research was carried out. AA and RB would
like to acknowledge the hospitality of the National Center for Theoretical Sciences
(NCTS), Physics Division in Taiwan.
\vspace{-0.5cm}

%

\end{document}